\documentclass[twocolumn,amsmath,amssymb]{revtex4-1}
\usepackage{color}
\usepackage{amstext,amssymb,amsfonts}
\usepackage{bm}
\usepackage{amsmath}
\usepackage{graphicx}
\usepackage{graphics}
\usepackage{epsfig}
\usepackage{graphicx,latexsym,amsmath,amssymb,amsfonts}
\usepackage{hyperref}

\newcommand{\bra}[1]{\ensuremath{\langle #1 \vert}}
\newcommand{\ket}[1]{\ensuremath{\vert #1  \rangle}}

\renewcommand{\b}[1]{\ensuremath{\mathbf{#1}}}
\renewcommand{\H}{\ensuremath{\text{H}}}

\renewcommand{\d}{\ensuremath{\text{d}}}
\newcommand{\lr}{\ensuremath{\text{lr}}}
\newcommand{\sr}{\ensuremath{\text{sr}}}
\newcommand{\s}{\ensuremath{\text{s}}}
\newcommand{\ee}{\ensuremath{\text{ee}}}

\newcommand{\HF}{\ensuremath{\text{HF}}}
\newcommand{\EXX}{\ensuremath{\text{EXX}}}

\DeclareMathOperator{\erf}{erf}
\DeclareMathOperator{\erfc}{erfc}

\begin{document}
\title{Linear-response range-separated density-functional theory \\ for atomic photoexcitation and photoionization spectra}
\author{Felipe Zapata} \email{felipe.zapata@lct.jussieu.fr}
\author{Eleonora Luppi} \email{eleonora.luppi@lct.jussieu.fr}
\author{Julien Toulouse} \email{toulouse@lct.jussieu.fr}
\affiliation{Laboratoire de Chimie Th\'eorique (LCT), Sorbonne Universit\'e and CNRS, F-75005 Paris, France}
\date{March 14, 2019}

\begin{abstract}
We investigate the performance of the range-separated hybrid (RSH) scheme, which combines long-range Hartree-Fock (HF) and a short-range density-functional approximation (DFA), for calculating photoexcitation/photoionization spectra of the H and He atoms, using a B-spline basis set in order to correctly describe the continuum part of the spectra. The study of these simple systems allows us to quantify the influence on the spectra of the errors coming from the short-range exchange-correlation DFA and from the missing long-range correlation in the RSH scheme. We study the differences between using the long-range HF exchange (nonlocal) potential and the long-range exact exchange (local) potential. Contrary to the former, the latter supports a series of Rydberg states and gives reasonable photoexcitation/photoionization spectra, even without applying linear-response theory. The most accurate spectra are obtained with the linear-response time-dependent range-separated hybrid (TDRSH) scheme. In particular, for the He atom at the optimal value of the range-separation parameter, TDRSH gives slightly more accurate photoexcitation and photoionization spectra than standard linear-response time-dependent HF. More generally, the present work shows the potential of range-separated density-functional theory for calculating linear and nonlinear optical properties involving continuum states.
\end{abstract}

\maketitle

\section{Introduction}

Nowadays, time-dependent density-functional theory (TDDFT)~\cite{RunGro-PRL-84}, applied within the linear-response formalism~\cite{GroKoh-PRL-85,Cas-INC-95,PetGosGro-PRL-96}, is a widely used approach for calculating photoexcitation spectra (transitions from bound to bound states) of electronic systems. In spite of many successes, it is however well known that usual (semi-)local density-functional approximations (DFAs), i.e. the local-density approximation (LDA) and generalized-gradient approximations (GGAs), for the exchange-correlation potential and its associated exchange-correlation kernel do not correctly describe long-range electronic transitions, such as those to Rydberg~\cite{CasJamCasSal-JCP-98} and charge-transfer~\cite{DreWeiHea-JCP-03} states in atomic and molecular systems. A better description of Rydberg excitations can be obtained with exchange-correlation potential approximations having the correct $-1/r$ long-range asymptotic decay~\cite{LeeBae-PRA-94,TozHan-JCP-98,CasSal-JCP-00,SchGriGisBae-JCP-00}, even though it has been shown that accurate Rydberg excitation energies and oscillator strengths can in fact be extracted from LDA calculations in small atoms~\cite{WasMaiBur-PRL-03,WasBur-PRL-05}. A more general solution for correcting both Rydberg and charge-transfer excitations is given by range-separated TDDFT approaches~\cite{TawTsuYanYanHir-JCP-04,YanTewHan-CPL-04,PeaHelSalKeaLutTozHan-PCCP-06,LivBae-PCCP-07,BaeLivSal-ARPC-10,FroKneJen-JCP-13,RebSavTou-MP-13} which express the long-range part of the exchange potential and kernel at the Hartree-Fock (HF) level. These range-separated approaches also give reasonably accurate values for the ionization energy threshold~\cite{YanTewHan-CPL-04,GerAng-CPL-05a,TsuSonSuzHir-JCP-10}.

Linear-response TDDFT has also been used for calculating photoionization spectra (transitions from bound to continuum states) of atoms and molecules~\cite{ZanSov-PRA-80,LevSov-PRA-84,SteDecLis-JPB-95,SteAltFroDec-CP-97,SteDec-JPB-97,SteDec-JCP-00,SteDecGor-JCP-01,SteFroDec-JCP-05,SteTofFroDec-JCP-06,TofSteDec-PRA-06,SteTofFroDec-TCA-07,ZhoChu-PRA-09}. These calculations are less standard in quantum chemistry since they involve spatial grid methods or B-spline basis sets for a proper description of the continuum states. In this case as well, usual (semi-)local DFAs provide a limited accuracy and asymptotically corrected exchange-correlation potential approximations give more satisfactory results. More accurate still, but less common, are photoionization spectra calculated with the exact-exchange (EXX) potential~\cite{SteDecGor-JCP-01} or the localized HF exchange potential and its associated kernel~\cite{ZhoChu-PRA-09}. Recently, range-separated approximations have been successfully used for calculating photoexcitation and photoionization spectra of molecular systems using time-propagation TDDFT with Gaussian basis sets together with an effective lifetime model compensating for the missing continuum states~\cite{LopGov-JCTC-13,FerBalLop-JCTC-15,SisAbaMauGaaSchLop-JCP-16}. However, to the best of our knowledge, range-separated approximations have not yet been used in frequency-domain linear-response TDDFT calculations of photoionization spectra.

In this work, we explore the performance of the linear-response time-dependent range-separated hybrid (TDRSH) scheme~\cite{RebSavTou-MP-13,TouRebGouDobSeaAng-JCP-13} for calculating photoexcitation and photoionization spectra of the H and He atoms using a B-spline basis set to accurately describe the continuum part of the spectra. The TDRSH scheme allows us to treat long-range exchange effects at the HF level and short-range exchange-correlation effects within (semi-)local DFAs. First, the dependence of the range-separated hybrid (RSH) orbital energies on the range-separation parameter is investigated, as well as the effect of replacing the long-range HF exchange nonlocal potential by the long-range EXX local potential (resulting in a scheme that we refer to as RSH-EXX). Second, oscillator strengths directly computed with the RSH and the RSH-EXX orbitals are compared with oscillator strengths obtained with the linear-response TDRSH scheme. The study of the H atom allows us to quantify the residual self-interaction error coming from the short-range exchange-correlation DFA, and the study of the He atom permits to quantify the effect of the missing long-range correlation in the RSH scheme. This work constitutes a first step for applying range-separated TDDFT to strong-field phenomena, such as high-harmonic generation or above-threshold ionization, where long-range effects and continuum states play an important role.

The outline of the paper is as follows. In Sec.~\ref{sec:theory}, firstly, we briefly review the RSH scheme and introduce the RSH-EXX variant, and, secondly, we review the linear-response TDRSH method. In Sec.~\ref{sec:implementation}, the basis set of B-spline functions is defined, and we indicate how the range-separated two-electron integrals are computed using an exact spherical harmonic expansion for the range-separated interaction. In Sec.~\ref{sec:results} results are presented and discussed. Firstly, we show the performance of the B-spline basis set for describing the density of continuum states of the H atom within the different methods. Secondly, the dependence of the orbital energies of the H and He atoms on the range-separation parameter is analyzed. Thirdly, different calculated photoexcitation/photoionization spectra for the H and He atoms are discussed and compared with exact results. In Sec.~\ref{sec:conclusions}, conclusions and perspectives are given. Unless otherwise indicated, Hartree atomic units are used throughout the paper.

\section{Range-separated density-functional theory}
\label{sec:theory}

\subsection{Range-separated hybrid scheme}

Range-separated density-functional theory (see, e.g., Refs.~\onlinecite{Sav-INC-96,TouColSav-PRA-04}) is based on the splitting of the Coulomb electron-electron interaction $w_\text{ee}(r)=1/r$ into long-range (lr) and short-range (sr) contributions
\begin{equation}
w_{\ee}(r)=w_{\ee}^\lr(r)+w_{\ee}^\sr(r),
\end{equation}
and the most common forms for the long-range and short-range interactions are
\begin{equation}
\label{erflr}
  w_{\ee}^\lr(r)=\frac{\erf(\mu r)}{r}, 
\end{equation}
and
\begin{equation}
\label{erfcsr}
  w_{\ee}^\sr(r)=\frac{\erfc(\mu r)}{r}.
\end{equation}
where $\erf$ and $\erfc$ are the error function and the complementary error function, respectively, and $\mu$ is a tunable range-separation parameter controlling the range of the separation. Using this decomposition, it is possible to rigorously combine a long-range wave-function approach with a complementary short-range DFA.

The simplest approach in range-separated density-functional theory consists in using a single-determinant wave function for the long-range interaction. This leads to the RSH scheme~\cite{AngGerSavTou-PRA-05} which spin orbitals $\{\varphi_{p}(\b{x})\}$ (where $\b{x}=(\b{r},\sigma)$ are space-spin coordinates) and orbital energies $\varepsilon_{p}$ can be determined for a given system by the following eigenvalue problem,
\begin{eqnarray}
\left( -\frac{1}{2} \bm{\nabla}^2 +v_{\text{ne}}(\b{r}) + v_\text{H}(\b{r}) + v_\text{xc}^{\sr}(\b{x}) \right) \varphi_{p}(\b{x}) 
\nonumber\\
+ \int v_{\text{x}}^{\lr,\HF}(\b{x},\b{x}') \varphi_{p}(\b{x}') \d \b{x}'  = \varepsilon_{p}\varphi_{p}(\b{x}),
\label{RSH}
\end{eqnarray}
where $v_{\text{ne}}(\b{r})$ is the nuclei-electron potential, $v_\text{H}(\b{r})$ is the Hartree potential for the Coulomb electron-electron interaction,
\begin{eqnarray}
v_\text{H}(\b{r}) = \int n(\b{x}') w_\ee(|\b{r}-\b{r}'|) \d \b{x}',
\end{eqnarray}
where $n(\b{x})=\sum_i^\text{occ} |\varphi_{i}(\b{x})|^2$ are the spin densities ($i$ refers to occupied spin orbitals), $v_{\text{x}}^{\lr,\HF}(\b{x},\b{x}')$ is the nonlocal HF exchange potential for the long-range electron-electron interaction,
\begin{eqnarray}
v_{\text{x}}^{\lr,\HF}(\b{x},\b{x}') = - \sum_{i}^\text{occ} \varphi_{i}^*(\b{x}') \varphi_{i}(\b{x}) w_\ee^\lr(|\b{r}-\b{r}'|),
\end{eqnarray}
and $v_\text{xc}^{\sr}(\b{x})$ is the short-range exchange-correlation potential
\begin{eqnarray}
v_\text{xc}^{\sr}(\b{x}) = \frac{\delta \bar{E}_\text{xc}^\sr}{\delta n(\b{x})},
\end{eqnarray}
where $\bar{E}_\text{xc}^\sr$ is the complement short-range exchange-correlation density functional. In this work, we use the short-range spin-dependent LDA exchange-correlation functional of Ref.~\onlinecite{PazMorGorBac-PRB-06} for $\bar{E}_\text{xc}^\sr$. The long-range and short-range potentials, $v_{\text{x}}^{\lr,\HF}(\b{x},\b{x}')$ and $v_\text{xc}^{\sr}(\b{x})$, explicitly depend on the range-separation parameter $\mu$, and consequently the spin orbitals, the orbital energies, and the density also implicitly depend on it. For $\mu=0$, $v_{\text{x}}^{\lr,\HF}(\b{x},\b{x}')$ vanishes and $v_\text{xc}^{\sr}(\b{x})$ becomes the usual full-range LDA exchange-correlation potential, and thus the RSH scheme reduces to standard Kohn-Sham LDA. For $\mu\to\infty$, $v_{\text{x}}^{\lr,\HF}(\b{x},\b{x}')$ becomes the usual full-range HF exchange potential and $v_\text{xc}^{\sr}(\b{x})$ vanishes, and thus the RSH scheme reduces to standard HF.

In the present paper, we also consider the following variant of the RSH scheme,
\begin{eqnarray}
\left( -\frac{1}{2} \bm{\nabla}^2 +v_{\text{ne}}(\b{r}) + v_\text{H}(\b{r}) + v_\text{xc}^{\sr}(\b{x}) +  v_\text{x}^{\lr,\EXX}(\b{x}) \right) \varphi_{p}(\b{x}) 
\nonumber\\
= \varepsilon_{p} \varphi_{p}(\b{x}), \;\;\;\;\;\;
\label{RSHEXX}
\end{eqnarray}
in which the long-range nonlocal HF exchange potential has been replaced by the long-range local EXX~\cite{TalSha-PRA-76,GorLev-PRA-94,GorLev-IJQC-95} potential
\begin{eqnarray}
\label{vexx}
v_\text{x}^{\lr,\EXX}(\b{x}) = \frac{\delta E_\text{x}^\lr}{\delta n(\b{x})},
\end{eqnarray}
where $E_\text{x}^\lr$ is the long-range exchange density functional~\cite{TouGorSav-IJQC-06,TouSav-JMS-06}. We will refer to this scheme as RSH-EXX. The calculation of the EXX potential is involved~\cite{FilUmrGon-PRA-96,Gor-PRL-99,IvaHirBar-PRL-99}, with the exception of one- and two-electron systems. Indeed, for one-electron systems, the long-range EXX potential is simply
\begin{eqnarray}
v_\text{x}^{\lr,\EXX}(\b{x}) = - v^{\lr}_\text{H}(\b{r}),
\end{eqnarray}
and for systems of two electrons in a single spatial orbital, it is
\begin{eqnarray}
v_\text{x}^{\lr,\EXX}(\b{x}) = - \frac{1}{2} v^{\lr}_\text{H}(\b{r}),
\end{eqnarray}
where $v^\lr_\text{H}(\b{r}) = \int n(\b{x}') w_\ee^\lr(|\b{r}-\b{r}'|) \d \b{x}'$ is the long-range Hartree potential. For these one- and two-electron cases, it can be shown that Eqs.~(\ref{RSH}) and~(\ref{RSHEXX}) give identical occupied orbitals but different unoccupied orbitals. More generally, for systems with more than two electrons, the HF and EXX exchange potentials give similar occupied orbitals but very different unoccupied orbitals.

Once orbitals and orbital energies are obtained from Eqs.~(\ref{RSH}) and~(\ref{RSHEXX}), the bare oscillator strengths can be calculated. They are defined as
\begin{eqnarray}
\label{oscillator0}
f_{ia}^0 = \frac{2}{3} \omega_{ia}^0 \sum_{\nu=x,y,z} |d_{\nu,ia}|^2,
\end{eqnarray}
where $i$ and $a$ refer to occupied and unoccupied spin orbitals, respectively, $\omega_{ia}^0 = \varepsilon_{a} -  \varepsilon_{i}$ are the bare excitation energies and $d_{\nu,ia} = \int \varphi_{i}^*(\b{x}) r_\nu  \varphi_{a}(\b{x}) \d\b{x}$ are the dipole-moment transition integrals. We will consider these bare excitation energies $\omega_{ia}^0$ and oscillator strengths $f_{ia}^0$ for a first approximation to photoexcitation/photoionization spectra.

\subsection{Linear-response time-dependent range-separated hybrid scheme}

In the time-dependent extension of the RSH scheme within linear response (referred to as TDRSH)~\cite{RebSavTou-MP-13,TouRebGouDobSeaAng-JCP-13,FroKneJen-JCP-13}, one has to solve the following pseudo-Hermitian eigenvalue equation
\begin{eqnarray}
\label{TDRSH}
\begin{pmatrix}
\b{A} & \b{B} \\
-\b{B}^* & -\b{A}^*
\end{pmatrix}
\begin{pmatrix}
\b{X}_n \\ \b{Y}_n
\end{pmatrix} = \omega_n
\begin{pmatrix}
\b{X}_n \\ \b{Y}_n
\end{pmatrix},
\end{eqnarray}
whose solutions come in pairs: excitation energies $\omega_n>0$ with eigenvectors $(\b{X}_n,\b{Y}_n)$, and de-excitation energies $\omega_n<0$ with eigenvectors $(\b{Y}_n^*,\b{X}_n^*)$. The elements of the matrices $\b{A}$ and $\b{B}$ are
\begin{eqnarray}
A_{ia,jb} = (\varepsilon_{a} -\varepsilon_{i}) \delta_{ij} \delta_{ab} + K_{ia,jb},
\end{eqnarray}
\begin{eqnarray}
B_{ia,jb} = K_{ia,bj},
\end{eqnarray}
where $i,j$ and $a,b$ refer to occupied and unoccupied RSH spin orbitals, respectively, and the coupling matrix $\b{K}$ contains the contributions from the Hartree kernel $f_\H(\b{r}_1,\b{r}_2)=w_\ee(|\b{r}_1-\b{r}_2|)$, the long-range HF exchange kernel $f_\text{x}^{\lr,\HF}(\b{x}_1,\b{x}_2;\b{x}_1',\b{x}_2')=-w_\ee^\lr(|\b{r}_1-\b{r}_2|) \delta(\b{x}_1-\b{x}_2') \delta(\b{x}_1'-\b{x}_2)$, and the adiabatic short-range exchange-correlation kernel $f_\text{xc}^{\sr}(\b{x}_1,\b{x}_2)=\delta v_\text{xc}^{\sr}(\b{x}_1)/\delta n(\b{x}_2)$
\begin{eqnarray}
\label{K}
K_{ia,jb} &=& \bra{aj} f_\H \ket{ib}+ \bra{aj} f_\text{x}^{\lr,\HF} \ket{ib} + \bra{aj} f_\text{xc}^{\sr} \ket{ib}
\nonumber\\
          &=& \bra{aj} w_\ee \ket{ib} - \bra{aj} w_\ee^\lr  \ket{bi} + \bra{aj} f_\text{xc}^{\sr} \ket{ib},
\end{eqnarray}
where $\bra{aj} w_\ee \ket{ib}$ and $\bra{aj} w_\ee^\lr  \ket{bi}$ are the two-electron integrals associated with the Coulomb and long-range interactions, respectively, and $\bra{aj} f_\text{xc}^{\sr} \ket{ib} = \iint \varphi_a^*(\b{x}_1) \varphi_j^*(\b{x}_2) f_\text{xc}^{\sr}(\b{x}_1,\b{x}_2) \varphi_i(\b{x}_1) \varphi_b(\b{x}_2) \d\b{x}_1 \d\b{x}_2 $. Since we use the short-range LDA exchange-correlation density functional, for $\mu=0$ the TDRSH scheme reduces to the usual linear-response time-dependent local-density approximation (TDLDA). For $\mu\to\infty$, the TDRSH scheme reduces to standard linear-response time-dependent Hartree-Fock (TDHF).

The time-dependent extension of the RSH-EXX variant within linear response (referred to as TDRSH-EXX) leads to identical equations with the exception that the long-range HF exchange kernel $f_\text{x}^{\lr,\HF}(\b{x}_1,\b{x}_2;\b{x}_1',\b{x}_2')$ is replaced by the long-range frequency-dependent EXX kernel~\cite{Gor-PRA-98,Gor-IJQC-98} $f_\text{x}^{\lr,\EXX}(\b{x}_1,\b{x}_2;\omega)=\delta v_\text{x}^{\lr,\EXX}(\b{x}_1,\omega)/\delta n(\b{x}_2,\omega)$. For one-electron systems, the long-range EXX kernel is simply
\begin{eqnarray}
f_\text{x}^{\lr,\EXX}(\b{x}_1,\b{x}_2;\omega)= -f_\H^\lr(\b{r}_1,\b{r}_2),
\end{eqnarray}
and, for systems with two electrons in a single spatial orbital, it is
\begin{eqnarray}
f_\text{x}^{\lr,\EXX}(\b{x}_1,\b{x}_2;\omega)= -\frac{1}{2} f_\H^\lr(\b{r}_1,\b{r}_2),
\end{eqnarray}
where $f_\H^\lr(\b{r}_1,\b{r}_2)=w_\ee^\lr(|\b{r}_1-\b{r}_2|)$ is the long-range Hartree kernel. For these one- and two-electron cases, TDRSH and TDRSH-EXX give rise to identical excitation energies and oscillator strengths.

Finally, we can calculate the corresponding TDRSH (or TDRSH-EXX) oscillator strengths as
\begin{eqnarray}
\label{oscillator}
f_{n} = \frac{2}{3} \omega_{n} \sum_{\nu=x,y,z} \left| d_{\nu,ia} (X_{n,ia} + Y_{n,ia}) \right|^2.
\end{eqnarray}
In the limit of a complete basis set, the linear-response oscillator strengths in Eq.~(\ref{oscillator}) always fulfill the Thomas-Reiche-Kuhn (TRK) sum rule, $\sum_n f_{n} = N$ where $N$ is the electron number. The bare oscillator strengths of Eq.~(\ref{oscillator0}) fulfill the TRK sum rule only in the case where the orbitals have been obtained from an effective local potential, i.e. for LDA and RSH-EXX but not for HF and RSH (see Ref.~\onlinecite{TouRebGouDobSeaAng-JCP-13}).

\section{Implementation in \\a B-spline basis set}
\label{sec:implementation}

In practice, each spin orbital is decomposed into a product of a spatial orbital and a spin function, $\varphi_{p}(\b{x})=\varphi_{p}(\b{r}) \delta_{\sigma_p,\sigma}$ where $\sigma_p$ is the spin of the spin orbital $p$, and we use spin-adapted equations. As we investigate atomic systems, the spatial orbitals are written in spherical coordinates,
\begin{equation}
 \varphi_{p}(\b{r})=R_{n_pl_p}(r)Y_{l_p}^{m_p}(\Omega),
\end{equation}
where $Y_{l_p}^{m_p}(\Omega)$ are the spherical harmonics ($\Omega$ stands for the angles $\theta,\phi$) and the radial functions $R_{n_pl_p}(r)$ are expressed as linear combinations of B-spline functions of order $k_\s$,
\begin{eqnarray}
 R_{n_p l_p}(r)=\sum_{\alpha=1}^{N_\s}c_\alpha^{n_p l_p}\frac{B^{k_\s}_\alpha(r)}{r},
\end{eqnarray}
where $N_\s$ is the dimension of the basis. To completely define a basis of B-spline functions, a non-decreasing sequence of $N_\s+k_\s$ knot points (some knot points are possibly coincident) must be given~\cite{Boor-78}. The B-spline function $B^{k_\s}_\alpha(r)$ is non zero only on the supporting interval $[r_\alpha,r_{\alpha+k_\s}]$ (containing $k_\s+1$ consecutive knot points) and is a piecewise function composed of polynomials of degree $k_\s-1$ with continuous first $k_\s-m$ derivatives across each knot of multiplicity $m$. We have chosen the first and the last knots to be $k_\s$-fold degenerate, i.e. $r_1 = r_2 = \cdots = r_{k_\s} = R_{\text{min}}$ and $r_{{N_\s+1}} = r_{{N_\s+2}} = \cdots = r_{{N_\s+k_\s}}= R_{\text{max}}$, while the multiplicity of the other knots is unity. The spatial grid spacing was chosen to be constant in the whole radial space between two consecutive non-coincident points and is given by $\Delta r = R_{\text{max}}/(N_\s-k_\s+1)$. In the present work, the first and the last B-spline functions were removed from the calculation to ensure zero boundary conditions at $r=R_{\text{min}}$ and $r=R_{\text{max}}$. The results presented in this paper have been obtained using the following parameters: $k_\s=8$, $N_\s=200$, $R_{\text{min}}=0$, and $R_{\text{max}} = 100$ bohr. Moreover, we need to use only s and p$_z$ spherical harmonics.

Working with such a B-spline representation, one must compute matrix elements involving integrals over B-spline functions. The principle of the calculation of one-electron and two-electron integrals over B-spline functions are well described by Bachau \emph{et al.} in Ref.~\onlinecite{BachCorDecHanMart-RepProgPhys-01}. We will now briefly review the computation of the standard Coulomb two-electron integrals over B-spline functions, and then we will present the calculation of the long-range or short-range two-electron integrals over B-spline functions, the latter being original to the present work.

\subsection{Coulomb two-electron integrals}

The Coulomb electron-electron interaction is given by
\begin{equation}
\label{coulomb}
 w_\ee(|\b{r}-\b{r}'|)=\frac{1}{\left(|\b{r}|^2+|\b{r}'|^2-2|\b{r}||\b{r}'|\cos\gamma \right)^{1/2}},
\end{equation}
where $\b{r}$ and $\b{r}'$ are electron vector positions and $\gamma$ is the angle between them. The multipolar expansion for this interaction is
\begin{equation}
\label{coulomb}
 w_\ee(|\b{r}-\b{r}'|)=\sum_{k=0}^{\infty}\left[\frac{r_<^k}{r_>^{k+1}}\right]\sum_{m_k=-k}^{k}(-1)^{m_k} C_{-m_k}^k(\Omega)C_{m_k}^k(\Omega'),
\end{equation}
where $r_<=\mathrm{min}(|\b{r}|,|\b{r}'|)$ and $r_>=\mathrm{max}(|\b{r}|,|\b{r}'|)$ and $C_{m_k}^k(\Omega)=\left(4\pi/(2k+1)\right)^{1/2}Y_k^{m_k}(\Omega)$ are the renormalized spherical harmonics. The Coulomb two-electron integrals, in the spatial orbital basis, can then be expressed as the sum of products of radial integrals and angular factors
\begin{eqnarray}
\label{coulomb_integral}
\nonumber
 \langle pq|w_\ee|tu\rangle&=&\sum_{k=0}^{\infty}R^k(p, q; t, u)\sum_{m_k=-k}^{k}\delta_{m_k,m_p-m_t}\delta_{m_k,m_q-m_u}\\
 &\times&(-1)^{m_k} c^k(l_p, m_p, l_t, m_t) c^k(l_q, m_q, l_u, m_u),
\end{eqnarray}
where $R^k(p, q; t, u)$ are the two-dimensional radial Slater integrals and the angular coefficients $c^k(l_p, m_p, l_t, m_t)$ and $c^k(l_q, m_q, l_u, m_u)$ are obtained from the Gaunt coefficients~\cite{RCowan-81,Cer-THESIS-12}. The coefficient $c^k(l, m, l', m')$ is non zero only if $|l-l'|\leq k \leq l+l'$ and if $l+l'+k$ is an even integer, which makes the sum over $k$ in Eq.~(\ref{coulomb_integral}) exactly terminate.
The Slater integrals are defined as
\begin{eqnarray}
\nonumber
 R^k(p, q; t, u)&=&\sum_{\alpha=1}^{N_\s}\sum_{\lambda=1}^{N_\s}\sum_{\beta=1}^{N_\s}\sum_{\nu=1}^{N_\s}c_{\alpha}^{n_pl_p}c_{\lambda}^{n_ql_q}c_{\beta}^{n_tl_t}c_{\nu}^{n_ul_u}\\
 & &\times R^k(\alpha, \lambda; \beta, \nu),
\end{eqnarray}
where $R^k(\alpha, \lambda; \beta, \nu)$ are the Slater matrix elements given by 
\begin{eqnarray}
\label{slaterelement}
\nonumber
 R^k(\alpha, \lambda; \beta, \nu)&=&\int_{0}^{\infty}\int_{0}^{\infty}B_{\alpha}^{k_\s}(r)B_{\lambda}^{k_\s}(r')\left[\frac{r_<^k}{r_>^{k+1}}\right]\\
 & &\times B_{\beta}^{k_\s}(r)B_{\nu}^{k_\s}(r')\d r \d r'.
\end{eqnarray}

In order to compute the Slater matrix elements $R^k(\alpha, \lambda; \beta, \nu)$, we have implemented the integration-cell algorithm developed by Qiu and Froese Fischer~\cite{CFFischer-99}. This algorithm exploits all possible symmetries and B-spline properties to evaluate efficiently the integrals in each two-dimensional radial region on which the integrals are defined. Gaussian quadrature is used to compute the integrals in each cell.  

\subsection{Long-range and short-range two-electron integrals}

A closed form of the multipolar expansion of the short-range electron-electron interaction defined in Eq.~(\ref{erfcsr}) was determined by \'Angy\'an \emph{et al.}~\cite{Janos-06}, following a previous work of Marshall~\cite{Marshall-02} who applied the Gegenbauer addition theorem to the Laplace transform of Eq.~(\ref{erfcsr}). This exact expansion is
\begin{eqnarray}
 w_{\ee}^{\sr}(|\b{r}-\b{r}'|)&=&\sum_{k=0}^{\infty}S^k(r_>,r_<;\mu)
\nonumber\\
&&\times\sum_{m_k=-k}^{k}(-1)^{m_k} C_{-m_k}^k(\Omega)C_{m_k}^k(\Omega'), \;\;\;
\end{eqnarray}
where the $\mu$-dependent radial function is written in terms of the scaled radial coordinates $\Xi=\mu \; r_>$ and $\xi=\mu\; r_<$ as
\begin{equation}
\label{srkernel}
 S^k(r_>,r_<;\mu)=\mu \; \Phi^k(\Xi,\xi),
\end{equation}
with
\begin{eqnarray}
\label{Phi}
\nonumber
 \Phi^k(\Xi,\xi)&=&H^k(\Xi,\xi)+F^k(\Xi,\xi)\\
 & &+\sum_{m=1}^k F^{k-m}(\Xi,\xi)\frac{\Xi^{2m}+\xi^{2m}}{(\xi\;\Xi)^m},
\end{eqnarray}
and the introduced auxiliary functions
\begin{eqnarray}
\nonumber
 H^k(\Xi,\xi)&=&\frac{1}{2(\xi\;\Xi)^{k+1}}\left[\left(\Xi^{2k+1}+\xi^{2k+1}\right)\erfc(\Xi+\xi)\right.\\
 & &\left.-\left(\Xi^{2k+1}-\xi^{2k+1}\right)\erfc(\Xi-\xi)\right], 
\end{eqnarray}
and 
\begin{eqnarray}
\nonumber
 F^k(\Xi,\xi)&=&\frac{2}{\pi^{1/2}}\sum_{p=0}^k\left(-\frac{1}{4(\xi\;\Xi)}\right)^{p+1}\frac{(k+p)!}{p!(k-p)!}\\
 & &\times\left[(-1)^{k-p} e^{-(\Xi+\xi)^2}-e^{-(\Xi-\xi)^2} \right].
\end{eqnarray}

In order to arrive at a separable expression in $\Xi$ and $\xi$, \'Angy\'an \emph{et al.}~\cite{Janos-06} also introduced a power series expansion of the radial function $\Phi^k(\Xi,\xi)$ in the smaller reduced variable $\xi$. However, the range of validity of this expansion truncated to the first few terms is limited to small values of $\xi$, i.e. $\xi \lesssim 1.5$, and higher-order expansions show spurious oscillations. After some tests, we decided to use the exact short-range radial function $\Phi^k(\Xi,\xi)$ without expansion in our work.

The expression of the short-range two-electron integrals $\langle pq|w^{\sr}_\ee|tu\rangle$ is then identical to the one in Eq.~(\ref{coulomb_integral}) with the simple difference that the radial term is not given by the standard Slater matrix elements. Now, the radial kernel in Eq.~(\ref{slaterelement}) is changed to that of Eq.~(\ref{srkernel}). Due to the fact that the radial kernel is not multiplicatively separable in the variables $r_>$ and $r_<$, the integration-cell algorithm is modified in order to calculate all integrals as non-separable two-dimensional integrals. In a second step, the long-range two-electron integrals can be simply obtained by difference
\begin{equation}
 \langle pq|w^{\lr}_\ee|tu\rangle=\langle pq|w_\ee|tu\rangle-\langle pq|w^{\sr}_\ee|tu\rangle.
\end{equation}

\section{Results and discussion}
\label{sec:results}

In this section, photoexcitation and photoionization spectra for the H and He atoms are presented. Photoexcitation and photoionization processes imply transitions from bound to bound and from bound to continuum states, respectively. For this reason, we first check the density of continuum states obtained with our B-spline basis set. After that, we show how orbital energies for the H and He atoms are influenced by the range-separation parameter $\mu$. Finally, having in mind these aspects, we discuss the different calculated spectra. All the studied transitions correspond to dipole-allowed spin-singlet transitions from the Lyman series, i.e. $1\text{s}\rightarrow n\text{p}$.

\subsection{Density of continuum states}
\label{sec:DOS}

\begin{figure}[t!]
\centering
\includegraphics[scale=0.4]{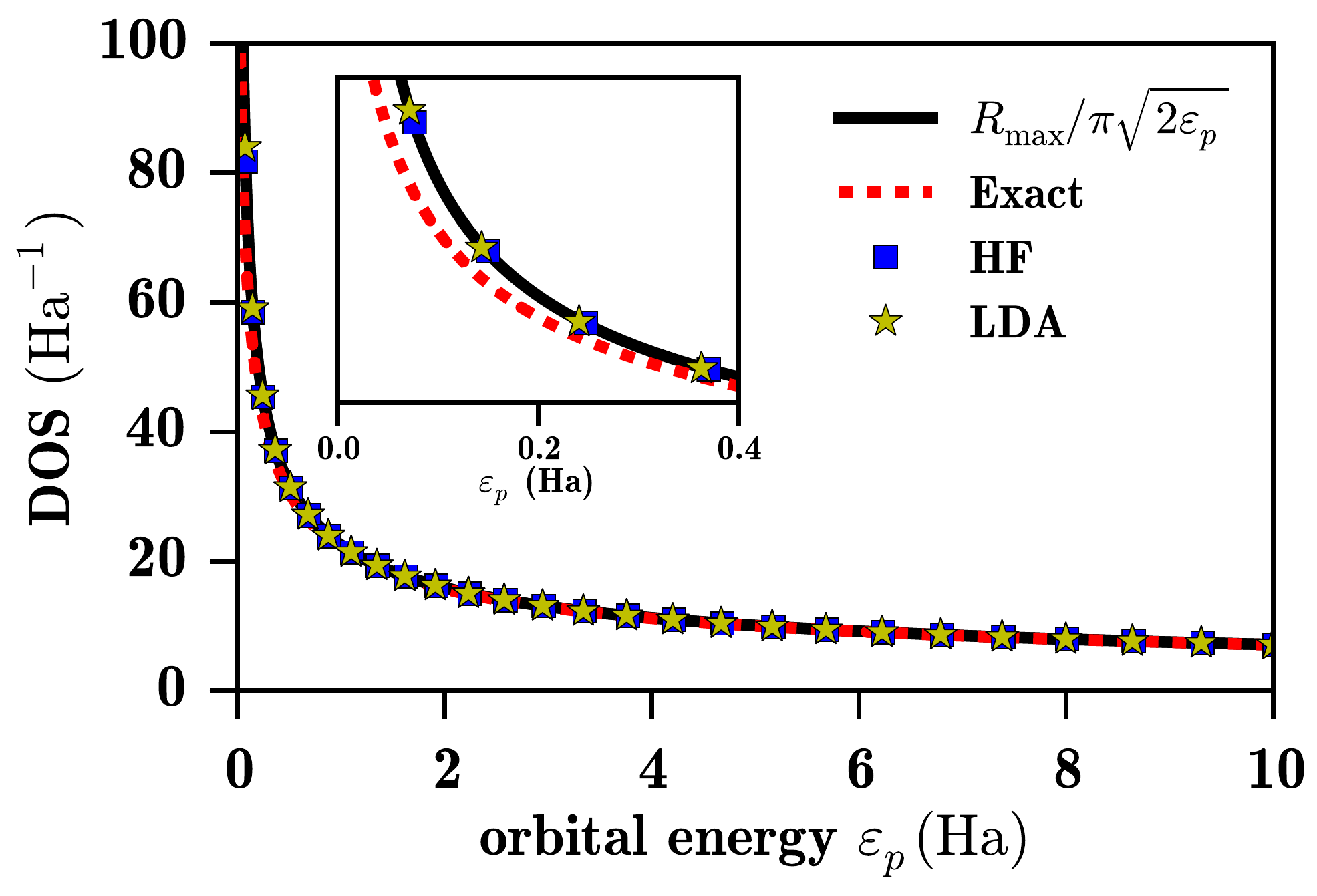}
\caption{Radial density of states (DOS) for a free particle, $\rho(\varepsilon_p)=R_\text{max}/\pi\sqrt{2\varepsilon_p}$, in a spherical box of size $R_\text{max} = 100$ bohr, and for the continuum p orbitals of the H atom computed with the exact Hamiltonian, or with the HF or LDA effective Hamiltonian using the B-spline basis set with the same $R_\text{max}$.}
\label{DOS}
\end{figure}

In Fig. \ref{DOS}, the radial density of states (DOS) of a free particle in a spherical box is compared with the radial DOS of the continuum p orbitals of the H atom computed with the exact Hamiltonian or with the HF or LDA effective Hamiltonian using the B-spline basis set. The radial DOS of a free particle is given by~\cite{BachCorDecHanMart-RepProgPhys-01} $\rho(\varepsilon)=R_\text{max}/\pi\sqrt{2\varepsilon}$ where $R_\text{max}$ is the radial size of the box, while for the different Hamiltonians using the B-spline basis set (with the same $R_\text{max}$) the radial DOS is calculated by finite differences as $\rho(\varepsilon_p)=2/(\varepsilon_{p+1} -\varepsilon_{p-1})$ where $\varepsilon_{p}$ are positive orbital energies.

As one can observe, the radial DOS computed with the LDA or the HF Hamiltonian is essentially identical to the DOS of the free particle. This can be explained by the fact that since the unoccupied LDA and HF orbitals do not see a $-1/r$ attractive potential they are all unbound and they all contribute to the continuum, similarly to the free-particle case. By contrast, for the exact Hamiltonian with the same B-spline basis set, one obtains a slightly smaller DOS in the low-energy region. This is due to the presence of the $-1/r$ attractive Coulomb potential which supports a series of bound Rydberg states, necessarily implying less unoccupied orbitals in the continuum for a given basis.

We have checked that, by increasing the size of the simulation box, together with the number of B-spline functions in the basis so as to keep constant the density of B-spline functions, the DOS of the exact Hamiltonian converges, albeit slowly, to the free-particle DOS. This must be the case since, for potentials vanishing at infinity, the global density of unbound states is independent of the potential for an infinite simulation box (only the local DOS depends on the potential, see e.g. Ref.~\onlinecite{Dic-BOOK-12}). From a numerical point of view, the computation of the DOS can be seen as a convergence test. With the present basis set, a huge energy range of the continuum spectrum is described correctly, and the difference between the DOS of the exact Hamiltonian and the free-particle DOS at low energies ($0.0 - 0.2$ Ha) is only about $10^{-4}$ Ha$^{-1}$. This difference is small enough to fairly compare the different methods considered in this paper.    

The calculation of the DOS is also important in order to compute proper oscillator strengths involving continuum states. Because of the use of a finite simulation box, the calculated positive-energy orbitals form, of course, a discrete set and not strictly a continuum. These positive-energy orbitals are thus not energy normalized as the exact continuum states should be. To better approximate pointwise the exact continuum wave functions, the obtained positive-energy orbitals should be renormalized. Following Mac\'ias \emph{et al.}~\cite{Macias88}, we renormalize the positive-energy orbitals by the square root of the DOS as $\tilde{\varphi}_p(\b{r})= \sqrt{\rho(\varepsilon_p)}\varphi_p(\b{r})$.

\subsection{Range-separated orbital energies}
\label{sec:orbital}

\begin{figure*}[t!]
\centering
\includegraphics[scale=0.4]{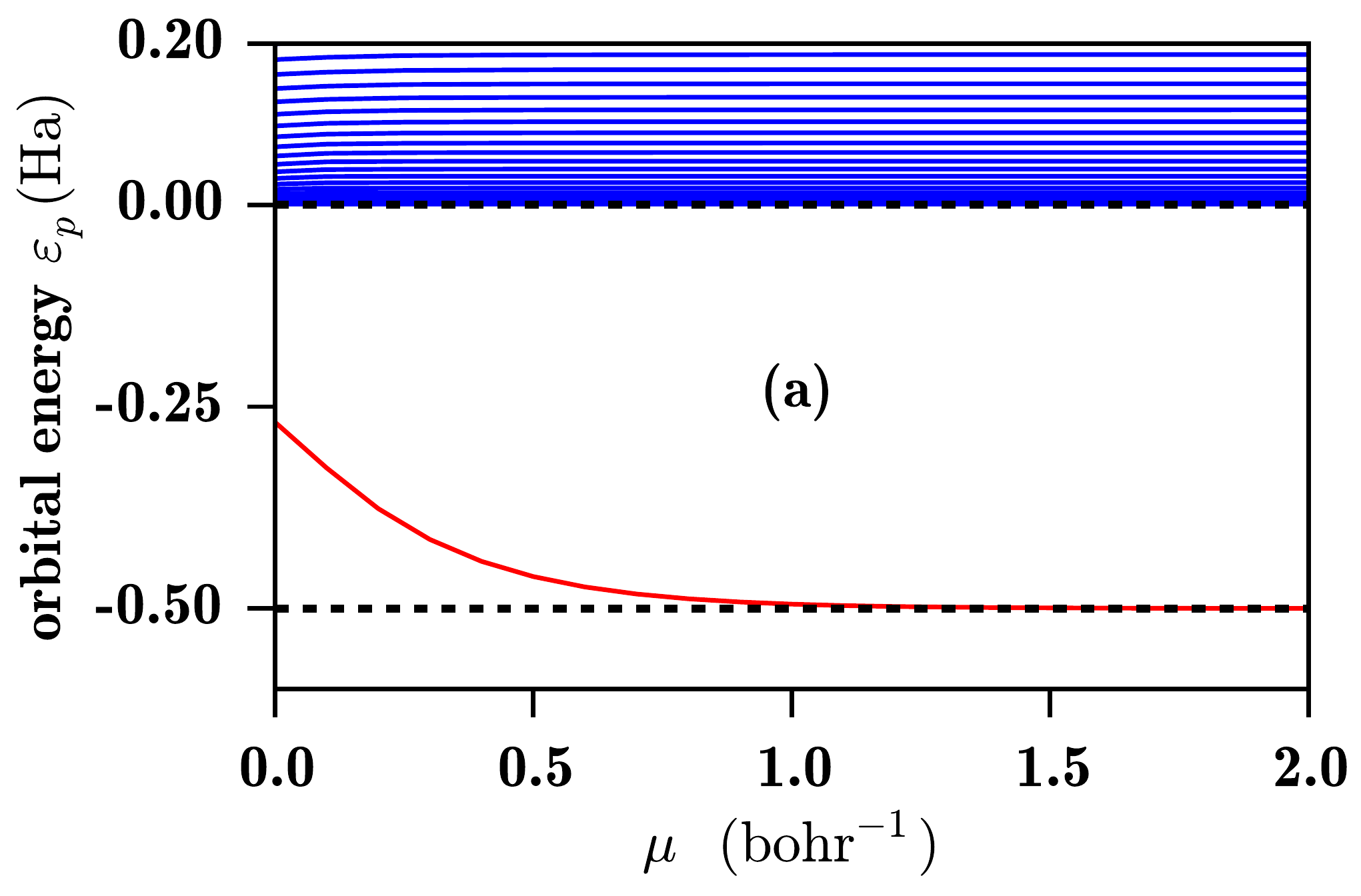} 
\includegraphics[scale=0.4]{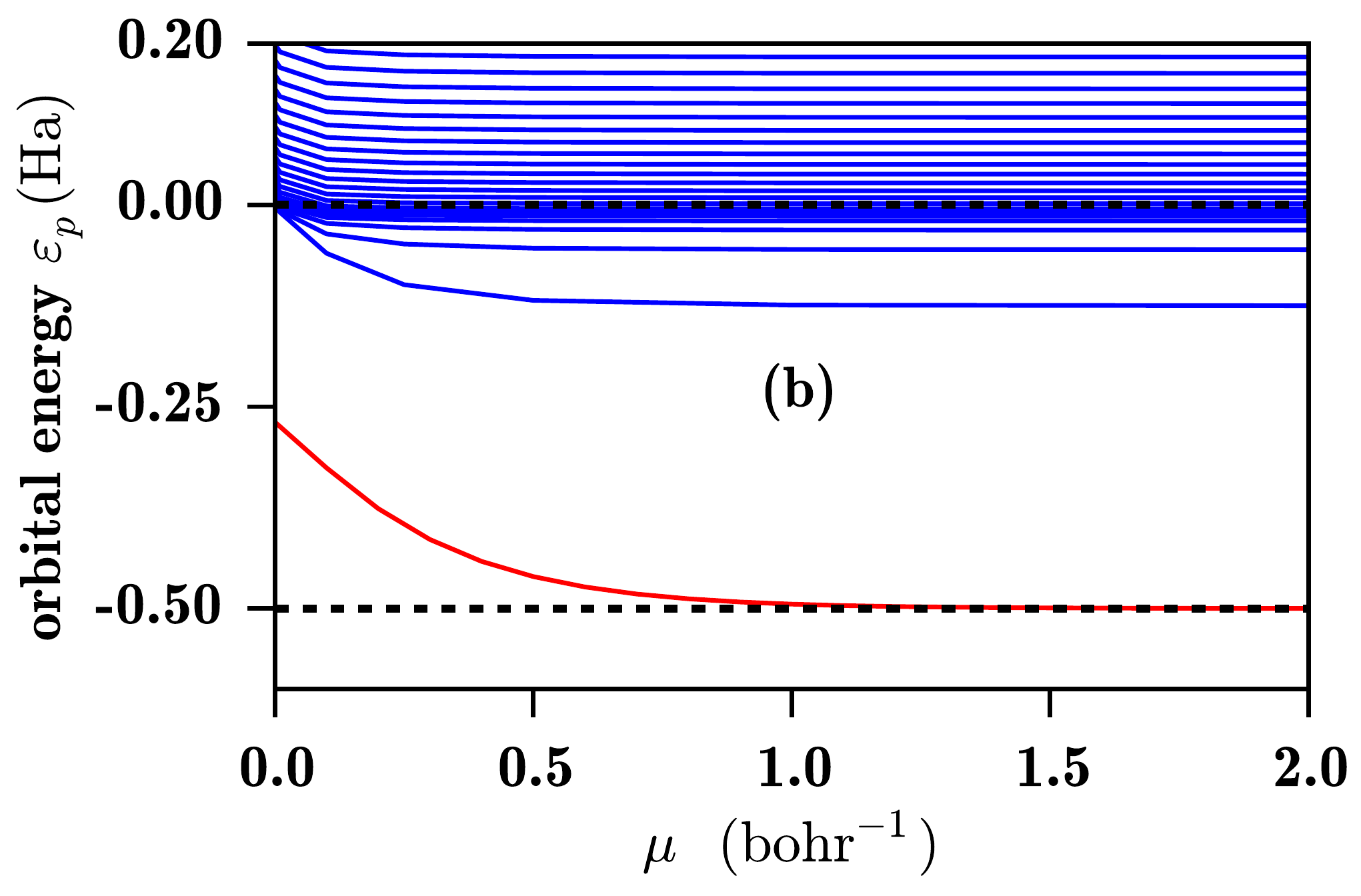} 
\caption{Orbital energies obtained with the RSH {\bf(a)} and with the RSH-EXX {\bf(b)} methods as a function of range-separation parameter $\mu$ for the H atom. The occupied 1s orbital energy is plotted in red and the unoccupied p orbital energies are plotted in blue. Horizontal dotted lines indicate the exact 1s orbital energy ($-0.5$ Ha) and the ionization limit (0 Ha).}
\label{Horbital}
\end{figure*}
\begin{figure*}[t!]
\centering
\includegraphics[scale=0.4]{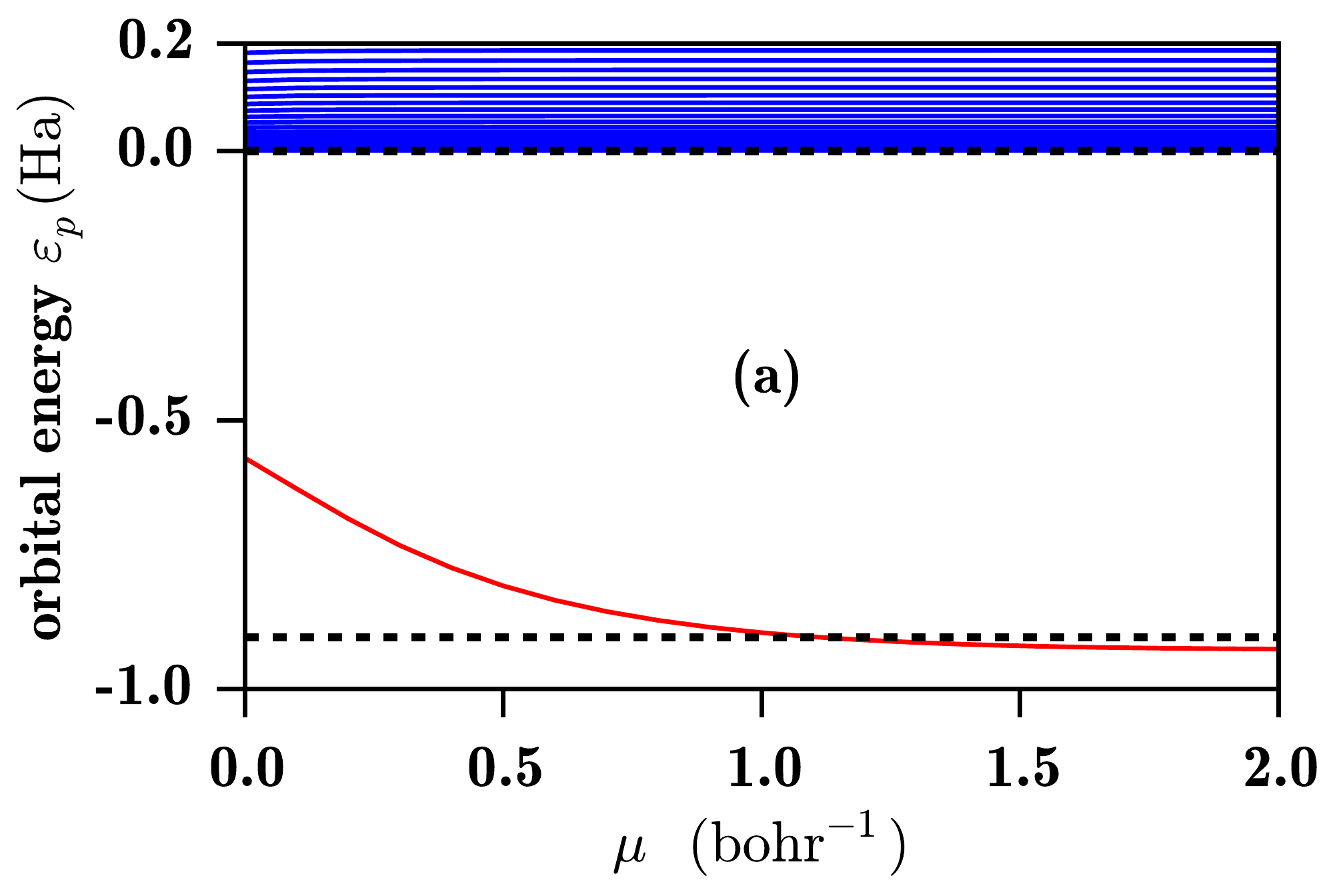} 
\includegraphics[scale=0.4]{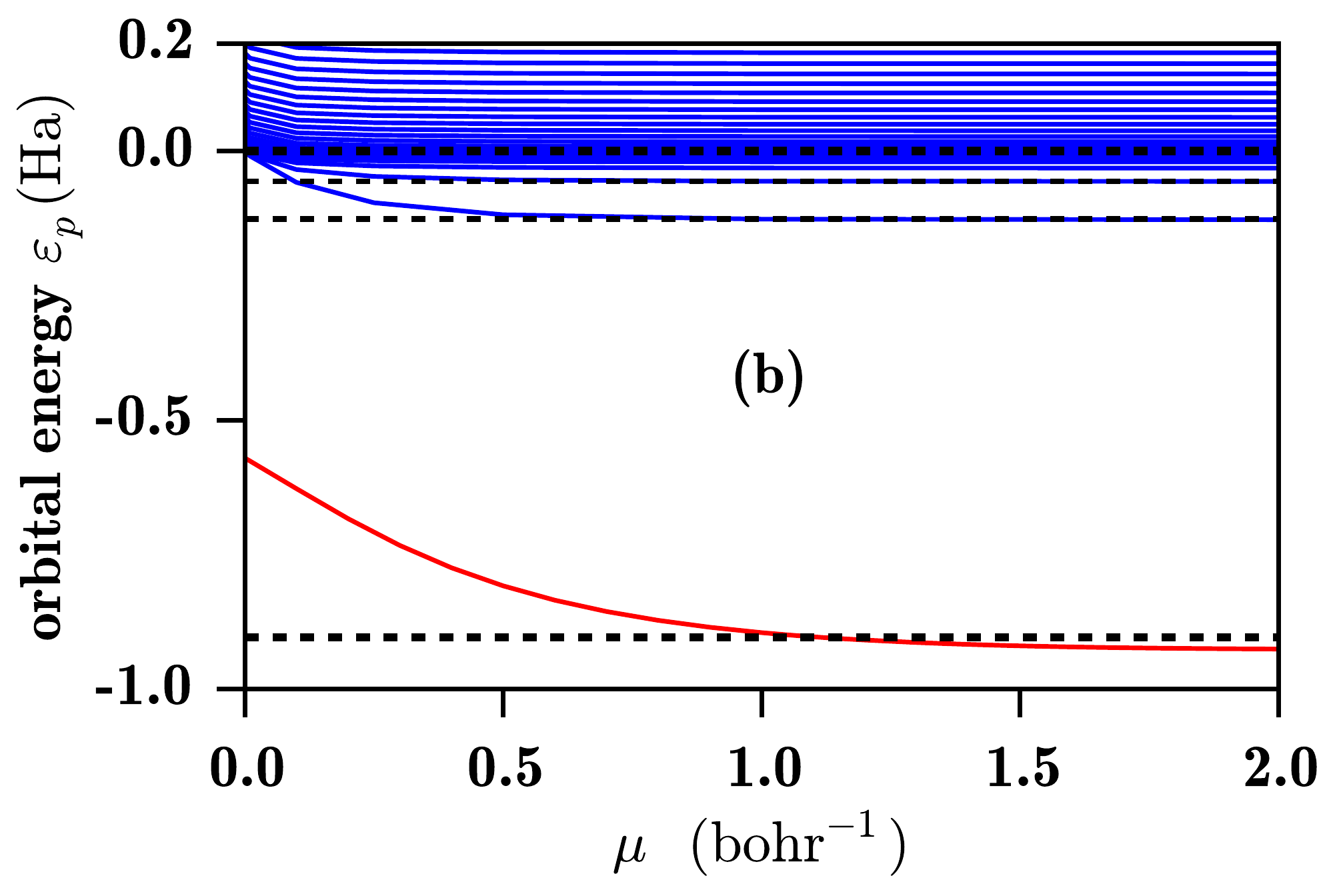} 
\caption{Orbital energies obtained with the RSH {\bf(a)} and with the RSH-EXX {\bf(b)} methods as a function of range-separation parameter $\mu$ for the He atom. The occupied 1s orbital energy is plotted in red and the unoccupied p orbital energies are plotted in blue. Horizontal dotted lines indicate exact Kohn-Sham orbital energies~\cite{UmrSavGon-INC-98}, including the opposite of the exact ionization energy ($-0.9036$ Ha) for the 1s orbital energy and the ionization limit (0 Ha).}
\label{Heorbital}
\end{figure*}

In Fig. \ref{Horbital} we show the 1s and the low-lying p orbital energies for the H atom calculated with both the RSH and RSH-EXX methods as a function of the range-separation parameter $\mu$. 

As one observes in Fig. \ref{Horbital}a, with the RSH method only the 1s ground state is bound, and the energy of this state is strongly dependent on $\mu$. At $\mu=0$, the self-interaction error introduced by the LDA exchange-correlation potential is maximal. But, when $\mu$ increases, the long-range HF exchange potential progressively replaces the long-range part of the LDA exchange-correlation potential and the self-interaction error is gradually eliminated until reaching the HF limit for $\mu\to\infty$, where one obtains the exact 1s orbital energy. The p orbitals (and all the other unoccupied orbitals) are always unbound and their (positive) energies are insensible to the value of $\mu$. One also observes that the approximate continuum of p orbitals has a DOS correctly decreasing as the energy increases, as previously seen in Fig. \ref{DOS}. 

In Fig. \ref{Horbital}b, one sees that the 1s orbital energy computed with the RSH-EXX method is identical to the 1s orbital energy obtained by the RSH scheme, as expected. However, a very different behavior is observed for the unoccupied p orbitals. Starting from the LDA limit at $\mu=0$ where all unoccupied orbitals are unbound, when the value of $\mu$ increases one sees the emergence of a series of bound Rydberg states coming down from the continuum. This is due to the introduction of an attractive $-1/r$ term in the long-range EXX potential, which supports a Rydberg series. For $\mu\to\infty$, we obtain the spectrum of the exact hydrogen Hamiltonian calculated with the B-spline basis set. Necessarily, with the finite basis used, the appearance of the discrete bound states is accompanied by a small reduction of the density of continuum states, as we already observed in Fig. \ref{DOS} with the exact Hamiltonian.

Another interesting aspect that can be observed in Fig. \ref{Horbital}b is the fact that the different bound-state energies reach their exact $\mu\to\infty$ values at different values of $\mu$. Thus, for a fixed small value of $\mu$, each bound-state energy is affected differently by the self-interaction error. For the compact 1s orbital, the self-interaction error is eliminated for $\mu \gtrsim 1$ bohr$^{-1}$. For the more diffuse 2p Rydberg state, the self-interaction error is essentially eliminated with $\mu \gtrsim 0.5$ bohr$^{-1}$. When we continue to climb in the Rydberg series, the orbitals become more and more diffuse and the self-interaction error is eliminated from smaller and smaller values of $\mu$.

In Fig. \ref{Heorbital}, the 1s and low-lying p orbital energies for the He atom are shown. Again, for the RSH method, one sees in Fig. \ref{Heorbital}a that only the occupied 1s orbital is bound and all the unoccupied p orbitals are in the continuum. Similarly to the case of the H atom, at $\mu=0$ the 1s orbital energy is too high, which can essentially be attributed to the self-interaction error in the LDA exchange-correlation potential. This error decreases when $\mu$ increases and the 1s orbital energy converges to its HF value for $\mu\to\infty$. However, contrary to the case of the H atom, for this two-electron system, the 1s HF orbital energy is not equal to the opposite of the exact ionization energy but is slightly too low due to missing correlation effects. In the spirit of the optimally tuned range-separated hybrids~\cite{LivBae-PCCP-07,SteKroBae-JACS-09,SteKroBae-JCP-09}, the range-separation parameter $\mu$ can be chosen so that the HOMO orbital energy is equal to the opposite of the exact ionization energy, which gives $\mu=1.115$ bohr$^{-1}$ for the He atom.

As regards the RSH-EXX method, one sees again in Fig. \ref{Heorbital}b that, for this two-electron system, the 1s RSH-EXX orbital energy is identical to the 1s RSH orbital energy. As in the case of the H atom, the introduction of the long-range EXX potential generates a series of bound  Rydberg states, whose energies converge to the Kohn-Sham EXX orbital energies for $\mu\to\infty$. For the Rydberg states of the He atom, it turns out that the Kohn-Sham EXX orbital energies are practically identical to the exact Kohn-Sham orbital energies~\cite{UmrSavGon-INC-98}, implying that the Kohn-Sham correlation potential has essentially no effect on these Rydberg states. As we will see, contrary to the RSH case, the set of unoccupied RSH-EXX orbitals can be considered as a reasonably good first approximation for the computation of photoexcitation and photoionization spectra, even before applying linear-response theory.

\subsection{Photoexcitation and photoionization \\ spectra for the hydrogen atom}

\begin{figure}[t!]
\centering
\includegraphics[scale=0.4]{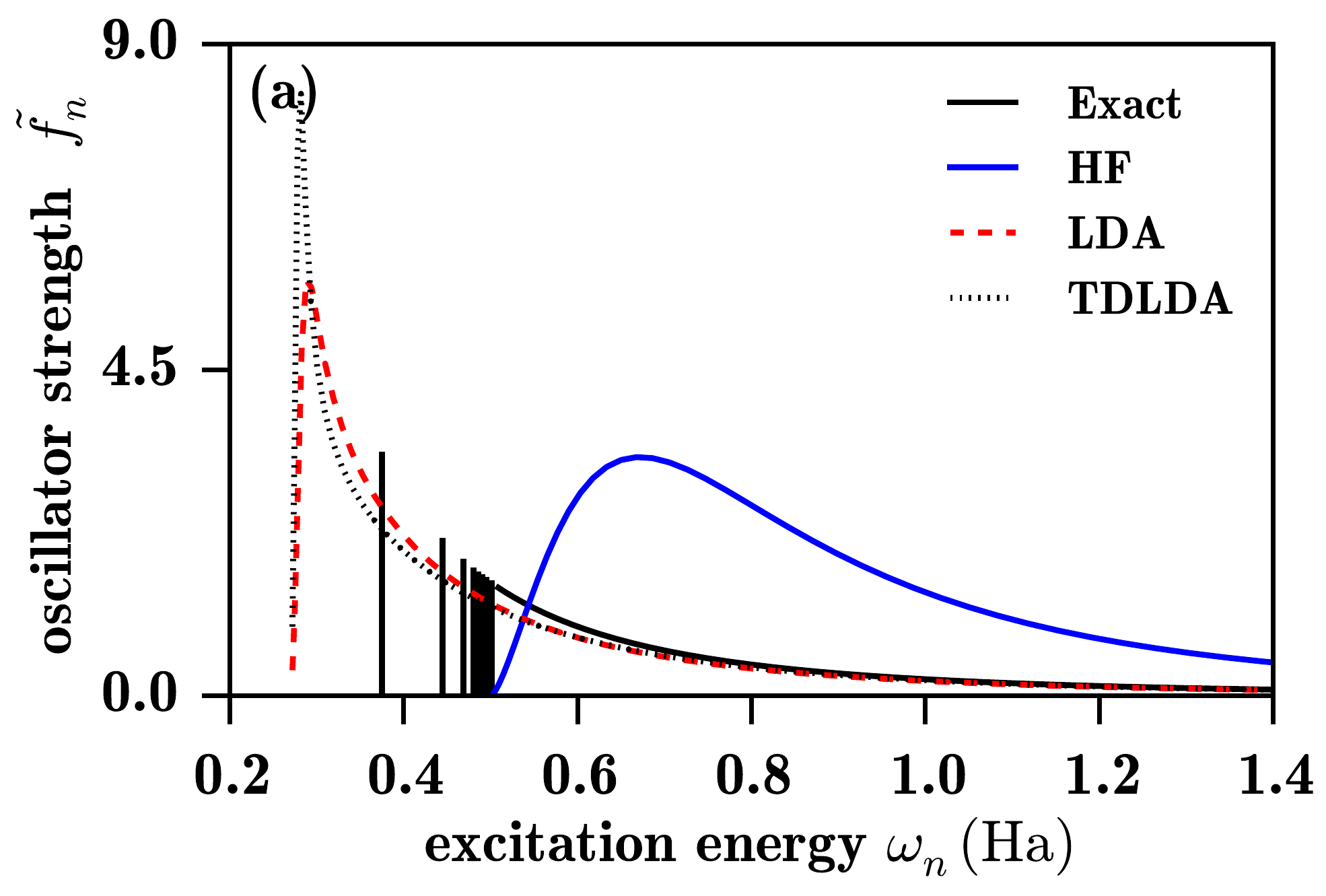}
\includegraphics[scale=0.4]{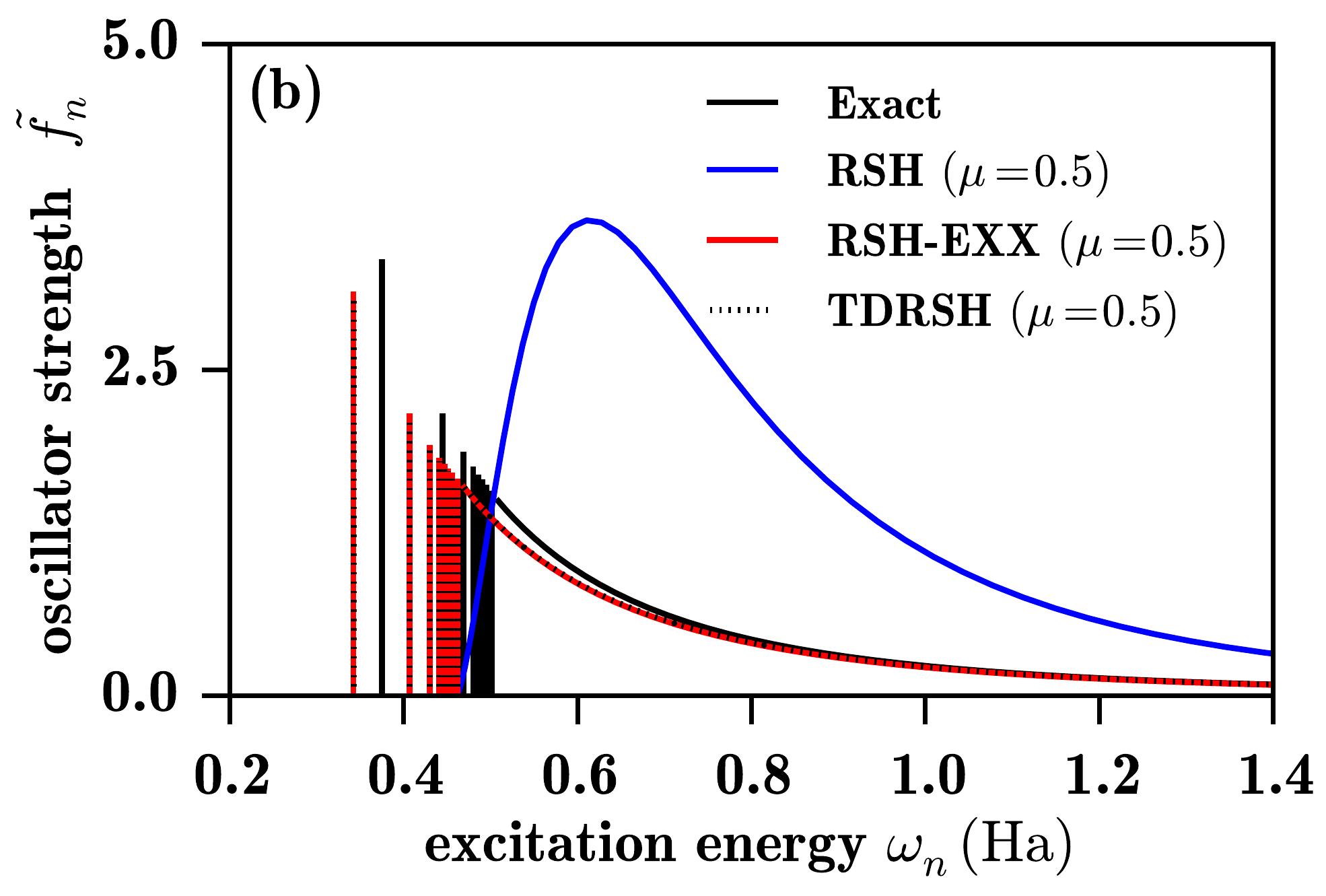}
\caption{Photoexcitation/photoionization spectra calculated with different methods for the H atom. In {\bf (a)} comparison of the HF, LDA, and TDLDA methods with respect to the calculation with the exact Hamiltonian. In {\bf (b)} comparison of the RSH, RSH-EXX, and TDRSH methods (all of them with a range-separation parameter of $\mu=0.5$ bohr$^{-1}$) with respect to the calculation with the exact Hamiltonian.}
\label{Hspectra}
\end{figure}

\begin{figure}[t!]
\centering
\includegraphics[scale=0.4]{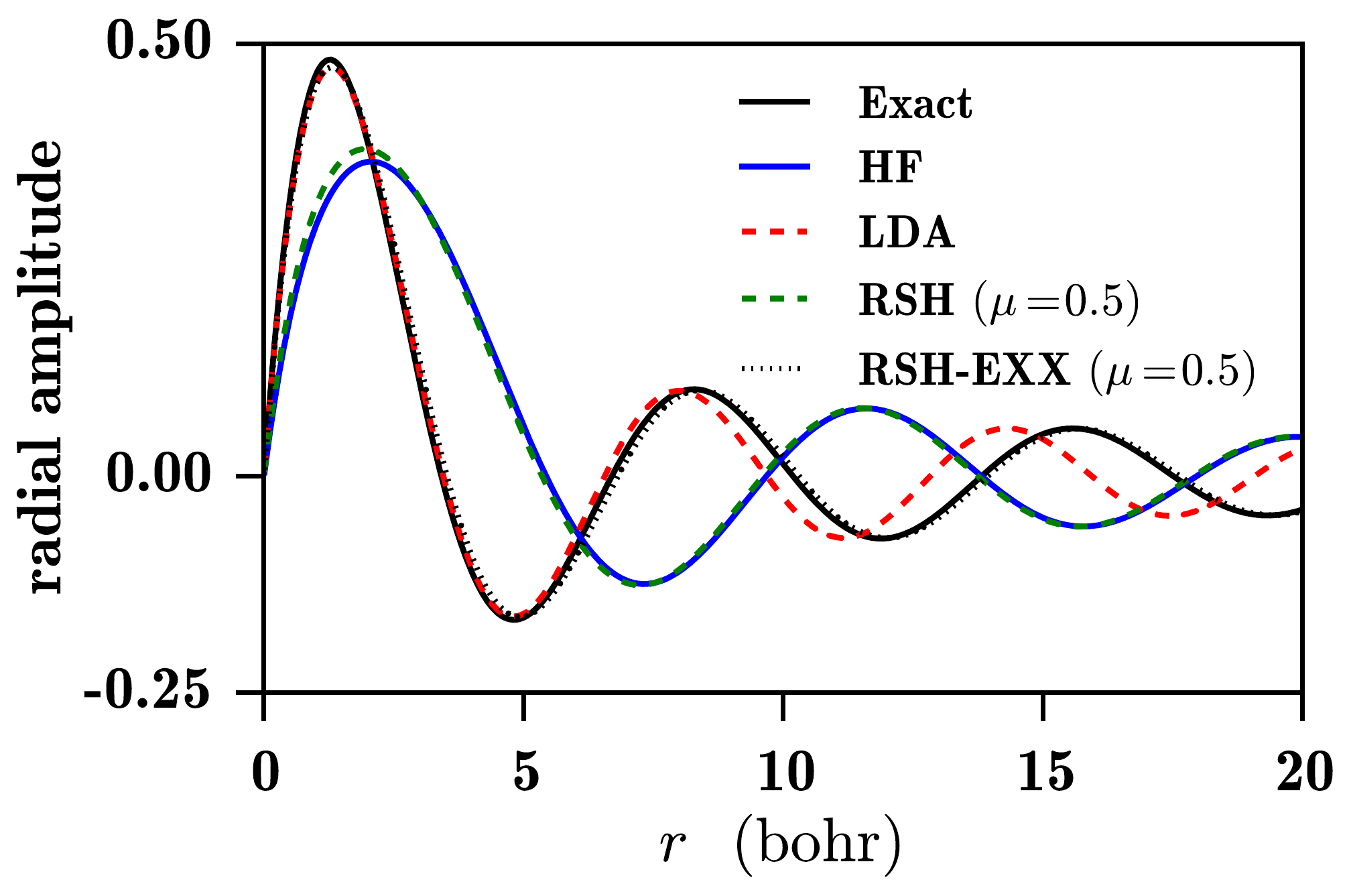}
\caption{Comparison of the renormalized radial amplitude $\tilde{R}(r) = \sqrt{\rho(\varepsilon)} R(r)$ of the continuum p orbital involved in the transition energy $\omega_n=\varepsilon - \varepsilon_\text{1s}=0.8$ Ha calculated 
by HF, LDA, RSH, and RSH-EXX (with a range-separation parameter of $\mu=0.5$ bohr$^{-1}$) with respect to the exact calculation for the H atom.}
\label{p-orbital}
\end{figure}

In Fig. \ref{Hspectra}, photoexcitation/photoionization spectra for the H atom calculated with different methods are shown. For the calculation using the exact Hamiltonian, the spectrum is correctly divided into a discrete and a continuum part, corresponding to the photoexcitation and photoionization processes, respectively. As already discussed in Sec.~\ref{sec:DOS}, for all calculations, the continuum states have been renormalized, or equivalently the oscillator strengths of the continuum part of the spectrum have been renormalized as $\tilde{f}_{1\text{s}\to n\text{p}} = \rho(\varepsilon_{n\text{p}}) f_{1\text{s}\to n\text{p}}$ where $\rho(\varepsilon_{n\text{p}})$ is the DOS at the corresponding positive orbital energy $\varepsilon_{n\text{p}}$. Moreover, for better readability of the spectra, following Refs.~\onlinecite{Friedrich98,WasMaiBur-PRL-03,YanFaaBur-JCP-09}, we have also renormalized the oscillator strengths of the discrete part of the spectrum as $\tilde{f}_{1\text{s}\to n\text{p}} = n^3 f_{1\text{s}\to n\text{p}}$ where $n$ is the principal quantum number of the excited p orbital. This makes the transition between the discrete and the continuum part of the spectrum smooth. Another thing is, since we are working with a finite B-spline basis set principally targeting a good continuum, we obtain only a limited number of Rydberg states and the last Rydberg states near the ionization threshold are not accurately described. In particular, the corresponding oscillator strengths are overestimated (not shown). To fix this problem, we could for example use quantum defect theory in order to accurately extract the series of Rydberg states~\cite{AlSharif98,Friedrich98,Faassen06,Faassen09}. However, for the propose of the present work, we did not find necessary to do that, and instead we have simply corrected the oscillator strengths of the last Rydberg states by interpolating between the oscillator strengths of the first five Rydberg states and the oscillator strength of the first continuum state using a second-order polynomial function of the type $\tilde{f}_n=c_0+c_1\;\omega_n+c_2\;\omega_n^2$. This procedure was applied for all spectra having a discrete part.

Let us first discuss the spectra in Fig. \ref{Hspectra}a. The LDA spectrum, calculated using the bare oscillator strengths of Eq.~(\ref{oscillator0}), does not possess a discrete photoexcitation part, which was of course expected since the LDA potential does not support bound Rydberg states, as seen in the $\mu=0$ limit of Fig.~\ref{Horbital}. The ionization threshold energy, giving the onset of the continuum spectrum, is much lower than the exact value (0.5 Ha) due to the self-interaction error in the ground-state orbital energy. At the ionization threshold, the LDA oscillator strengths are zero, in agreement with the Wigner-threshold law~\cite{Wig-PR-48,SadBohCavEsrFabMacRau-JPB-00} for potentials lacking a long-range attractive $-1/r$ Coulomb tail. Close above the ionization threshold, the LDA spectrum has an unphysical large peak, which corresponds to continuum states with an important local character. However, as noted in Ref.~\onlinecite{WasMaiBur-PRL-03}, at the exact Rydberg transition energies, the LDA continuum oscillator strengths are actually reasonably good approximations to the exact discrete oscillator strengths, which was explained by the fact that the LDA potential is approximately the exact Kohn-Sham potential shifted by a constant. Moreover, above the exact ionization energy, LDA reproduces relatively well the exact photoionization spectrum and becomes essentially asymptotically exact in the high-energy limit. This is consistent with the fact that, at a sufficiently high transition energy, the LDA continuum orbitals are very similar to the exact ones, at least in the spatial region relevant for the calculation of the oscillation strengths, as shown in Fig. \ref{p-orbital}.

The TDLDA spectrum differs notably from the LDA spectrum only in that the unphysical peak at around $0.3$ Ha, close above its ionization threshold, has an even larger intensity. This increased intensity comes from the contribution of the LDA exchange-correlation kernel (not shown). The LDA exchange-correlation kernel being local, its larger impact is for the low-lying LDA continuum orbitals having a local character. As the TRK sum rule must be satisfied, the higher peak in the TDLDA spectrum is followed by a decrease of the oscillator strengths faster than in the LDA spectrum, until they reach the same asymptotic behavior.

The HF spectrum in Fig. \ref{Hspectra}a not only has no discrete photoexcitation part, as expected since the unoccupied HF orbitals are unbound (see the $\mu\to\infty$ limit of Fig.~\ref{Horbital}a), but does not even look as a photoionization spectrum. The HF unoccupied orbitals actually represent approximations to the continuum states of the H$^-$ anion, and are thus much more diffuse than the exact continuum states of the H atom, as shown in Fig. \ref{p-orbital}. Consequently, the HF spectrum has in fact the characteristic shape of the photodetachment spectrum of the H$^-$ anion~\cite{BetSal-BOOK-57,Rau-JAA-96} (with the caveat that the initial state is the 1s orbital of the H atom instead of the 1s orbital of the H$^-$ anion). Finally, note that, for the H atom, linear-response TDHF gives of course the exact photoexcitation/photoionization spectrum.

Let us now discuss the spectra obtained with the range-separated methods in Fig. \ref{Hspectra}b. The common value of the range-separation parameter $\mu=0.5$ bohr$^{-1}$ has been used~\cite{GerAng-CPL-05a}. The RSH spectrum looks like the photodetachment spectrum of the H$^-$ anion. This is not surprising since the RSH effective Hamiltonian contains a long-range HF exchange potential. The RSH continuum orbitals are similarly diffuse as the HF continuum orbitals, as shown in Fig.~\ref{p-orbital}. The RSH ionization threshold energy is slightly smaller than the exact value (0.5 Ha) due to the remaining self-interaction error in the 1s orbital energy stemming from the short-range LDA exchange-correlation potential at this value of $\mu$. The RSH-EXX ionization threshold is identical to the RSH one, but, contrary to the RSH spectrum, the RSH-EXX spectrum correctly shows a discrete photoexcitation part and a continuum photoionization part. Beside the small redshift of the spectrum, the self-interaction error at this value of $\mu$ manifests itself in slightly too small RSH-EXX oscillator strengths. The RSH-EXX continuum orbitals are very similar to the exact continuum orbitals, as shown in Fig.~\ref{p-orbital}. Finally, at this value of $\mu$, TDRSH gives a photoexcitation/photoionization spectrum essentially identical to the RSH-EXX spectrum.

\subsection{Photoexcitation and photoionization \\ spectra for the helium atom}

\begin{figure}[t!]
\centering
\includegraphics[scale=0.4]{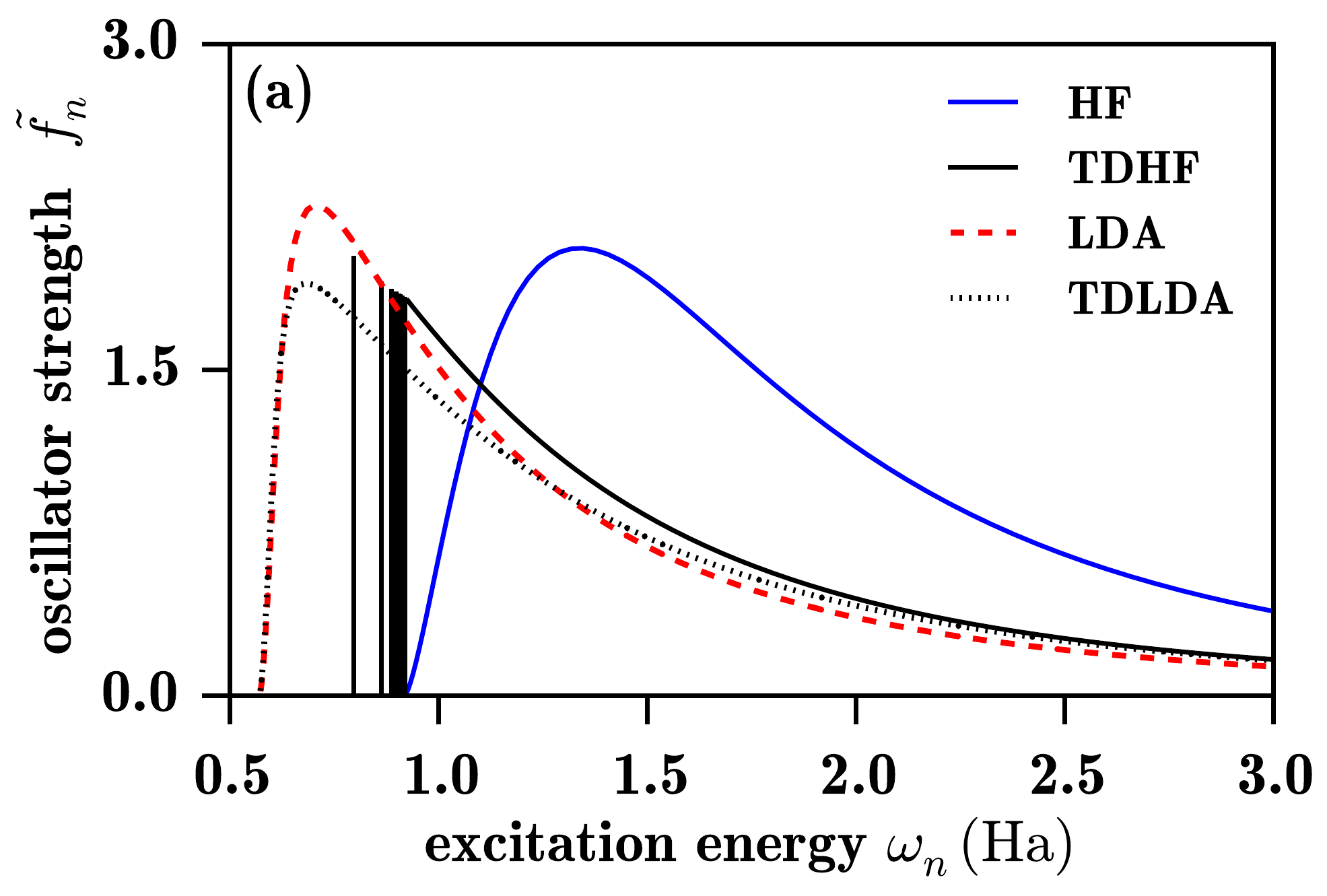}
\includegraphics[scale=0.4]{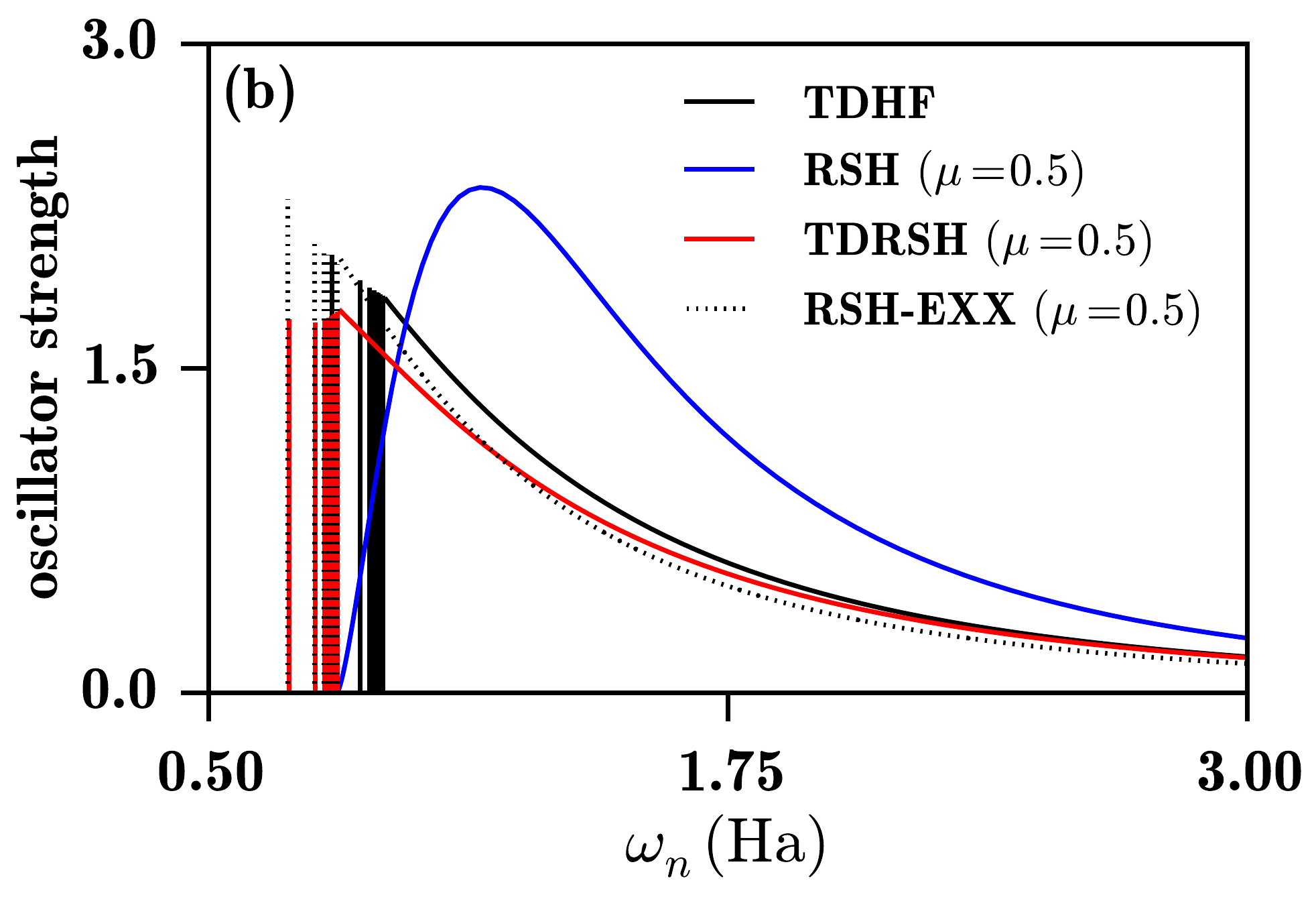}
\caption{Photoexcitation and photoionization spectra calculated with different methods for the He atom. In {\bf (a)} comparison of HF, TDHF, LDA, and TDLDA methods. In {\bf (b)} comparison of RSH, RSH-EXX, and TDRSH methods (all of them with a range-separation parameter of $\mu=1.115$ bohr$^{-1}$).}
\label{Hespectra}
\end{figure}
\begin{figure}[h!]
\centering
\includegraphics[scale=0.4]{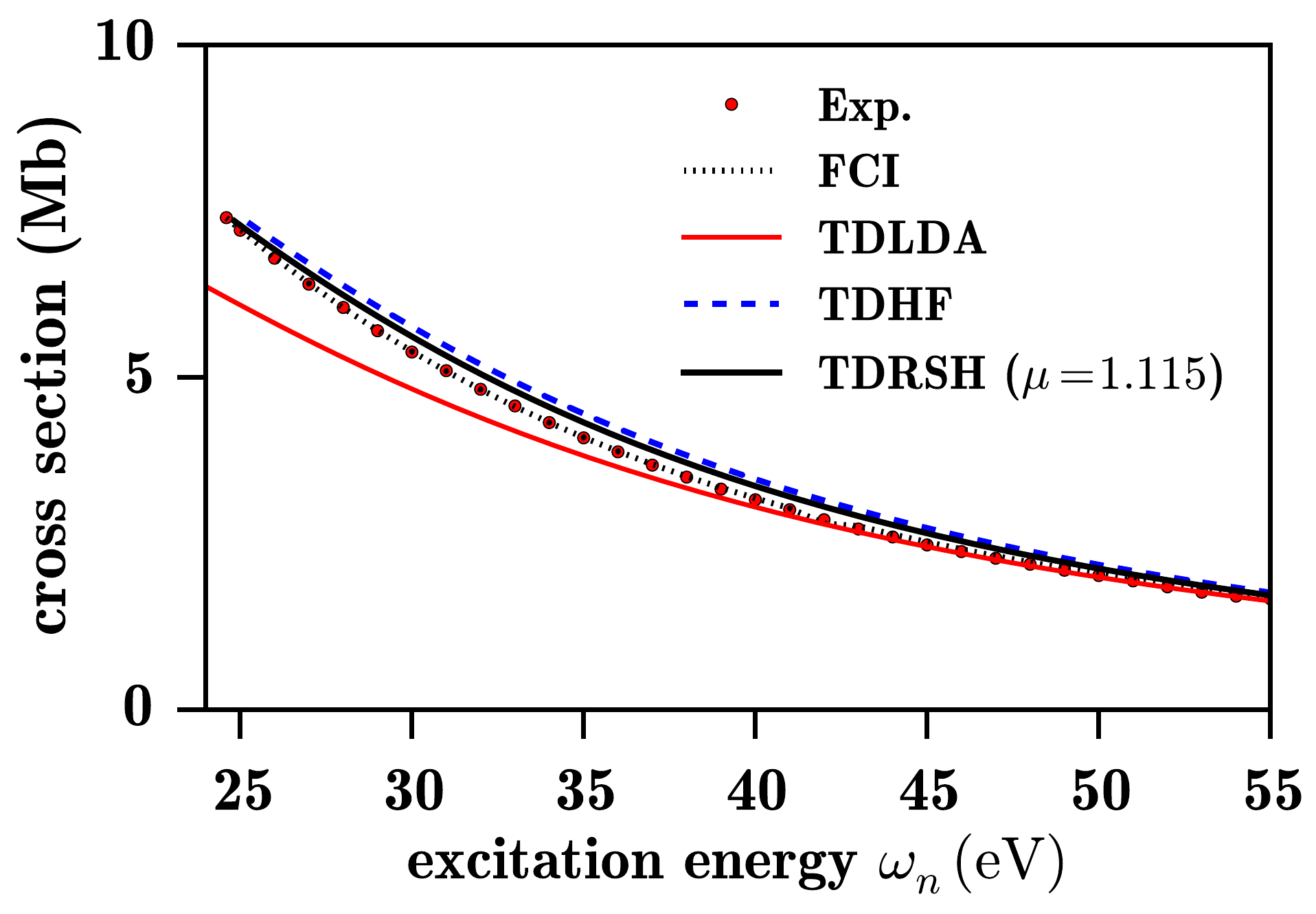}
\caption{Photoionization cross-section profile for the He atom. Normalized cross sections are given (in Hartree atomic units) by $\sigma_{n}= (2\pi^2/c) \tilde{f}_{n}$ where $\tilde{f}_{n}$ are the renormalized oscillator strengths and $c$ is the speed of light. Conversion factors 1 Ha = 27.207696 eV and 1 bohr$^2=28.00283$ Mb are employed. The experimental data and the FCI results are from Ref. \onlinecite{Venuti96}. }
\label{exp}
\end{figure}

\begin{table*}
\caption{Excitation energies ($\omega_n$ in Ha) and oscillator strengths ($f_n$) of the first discrete transitions calculated with different methods for the He atom. The ionization energy is also given.}
\begin{tabular}{c c c c c c c c c c c c}
\hline
\hline
Transition             & \multicolumn{2}{c}{\phantom{xxxxx}Exact$^a$\phantom{xxxxx}}   && \multicolumn{2}{c}{\phantom{xxxxx}TDHF\phantom{xxxxx}}   &&  \multicolumn{2}{c}{RSH-EXX ($\mu=1.115$)}  &&  \multicolumn{2}{c}{TDRSH ($\mu=1.115$)} \\
\cline{2-3}\cline{5-6}\cline{8-9}\cline{11-12}
                       & $\omega_n$ & $f_n$  && $\omega_n$ & $f_n$ && $\omega_n$ & $f_n$ && $\omega_n$ & $f_n$\\
\hline
1$^1$S $\to$ 2$^1$P         & 0.7799 & 0.2762 && 0.7970  & 0.2518  && 0.7766 & 0.3303   && 0.7827 & 0.2547   \\
1$^1$S $\to$ 3$^1$P         & 0.8486 & 0.0734 && 0.8636  & 0.0704  && 0.8474 & 0.0857   && 0.8493 & 0.0708   \\
1$^1$S $\to$ 4$^1$P         & 0.8727 & 0.0299 && 0.8872  & 0.0291  && 0.8721 & 0.0344   && 0.8729 & 0.0292  \\
1$^1$S $\to$ 5$^1$P         & 0.8838 & 0.0150 && 0.8982  & 0.0148  && 0.8835 & 0.0172   && 0.8839 & 0.0148  \\
1$^1$S $\to$ 6$^1$P         & 0.8899 & 0.0086 && 0.9042  & 0.0087  && 0.8897 & 0.0100   && 0.8899 & 0.0087  \\
\hline
Ionization energy           & \multicolumn{2}{c}{0.9036}  && \multicolumn{2}{c}{0.9180}  && \multicolumn{2}{c}{0.9036} && \multicolumn{2}{c}{0.9036}    \\
\hline
\hline
$^a$From Ref.~\onlinecite{KonHat-PRA-84}.
\end{tabular}
\label{tab:helium}
\end{table*}

In Fig. \ref{Hespectra}, different photoexcitation/photoionization spectra for the He atom are shown. As in the H atom case, the oscillator strengths of the discrete part of the TDHF, RSH-EXX, and TDRSH spectra have been interpolated (using again the oscillator strengths of first five Rydberg states and of the first continuum state) to correct the overestimation of the oscillator strengths for the last Rydberg transitions. The excitation energies and the (non-interpolated) oscillator strengths of the first five discrete transitions are reported in Table~\ref{tab:helium} and compared with exact results. The photoionization part of some of the calculated spectra are compared with full configuration-interaction (FCI) calculations and experimental results in Fig.~\ref{exp}.

In Fig. \ref{Hespectra}a, one sees that the HF spectrum looks again like a photodetachment spectrum, corresponding in this case to the He$^-$ anion. By contrast, TDHF gives a reasonable photoexcitation/photoionization spectrum. In particular, for the first discrete transitions listed in Table~\ref{tab:helium}, TDHF gives slightly too large excitation energies by at most about 0.02 Ha (or 0.5 eV) and slightly too small oscillator strengths by at most about 0.025. The ionization energy is also slightly too large by about 0.015 Ha, as already seen from the HF 1s orbital energy in the $\mu\to\infty$ limit of Fig.~\ref{Heorbital}. As regards the photoionization part of the spectrum, one sees in Fig.~\ref{exp} that TDHF gives slightly too large photoionization cross sections.

The LDA spectrum in Fig. \ref{Hespectra}a is also similar to the LDA spectrum for the H atom. The ionization threshold energy is much too low, and the spectrum lacks a discrete part and has an unphysical maximum close above the ionization threshold. Except from that, taking as reference the TDHF spectrum (which is close to the exact spectrum), the LDA spectrum is a reasonable approximation to the photoionization spectrum and, again as noted in Ref.~\onlinecite{WasMaiBur-PRL-03}, a reasonable continuous approximation to the photoexcitation spectrum. In comparison to LDA, TDLDA~\cite{ZapLupTou-JJJ-XX-note} gives smaller and less accurate oscillator strengths in the lower-energy part of the spectrum but, the TRK sum rule having to be preserved, larger oscillator strengths in the higher-energy part of the spectrum, resulting in an accurate high-energy asymptotic behavior as seen in Fig.~\ref{exp}.

Fig. \ref{Hespectra}b shows the spectra calculated with RSH, RSH-EXX, and TDRSH using for the range-separation parameter the value $\mu=1.115$ bohr$^{-1}$ which imposes the exact ionization energy, as explained in Sec.~\ref{sec:orbital}. The RSH spectrum is similar to the HF spectrum and does not represent a photoexcitation/photoionization spectrum. By contrast, the RSH-EXX spectra is qualitatively correct for a photoexcitation/photoionization spectrum. As shown in Table~\ref{tab:helium}, in comparison with TDHF, RSH-EXX gives more accurate Rydberg excitation energies, with a largest error of about 0.003 Ha (or 0.08 eV), but less accurate oscillator strengths which are significantly overestimated. The TDRSH method also gives a correct photoexcitation/photoionization spectrum, with the advantage that it gives Rydberg excitation energies as accurate as the RSH-EXX ones and corresponding oscillator strengths as accurate as the TDHF ones. As shown in Fig.~\ref{exp}, TDRSH also gives a slightly more accurate photoionization cross-section profile than TDHF.

\section{Conclusions}
\label{sec:conclusions}

We have investigated the performance of the RSH scheme for calculating photoexcitation/photoionization spectra of the H and He atoms, using a B-spline basis set in order to correctly describe the continuum part of the spectra. The study of these simple systems allowed us to quantify the influence on the spectra of the errors coming from the short-range exchange-correlation LDA and from the missing long-range correlation in the RSH scheme. For the He atom, it is possible to choose a value for the range-separation parameter $\mu$ for which these errors compensate each other so as to obtain the exact ionization energy.

We have studied the differences between using the long-range HF exchange nonlocal potential and the long-range EXX local potential. Contrary to the former, the latter supports a series of Rydberg states and the corresponding RSH-EXX scheme, even without applying linear-response theory, gives reasonable photoexcitation/photoionization spectra. Nevertheless, the most accurate spectra are obtained with linear-response TDRSH (or TDRSH-EXX since they are equivalent for one- and two-electron systems). In particular, for the He atom at the optimal value of $\mu$, TDRSH gives slightly more accurate photoexcitation and photoionization spectra than standard TDHF.

The present work calls for further developments. First, the merits of TDRSH (and/or TDRSH-EXX) for calculating photoexcitation/photoionization spectra of larger atoms and molecules, where screening effects are important, should now be investigated. Second, it would be interesting to test the effects of going beyond the LDA for the short-range exchange-correlation functional~\cite{TouColSav-JCP-05,GolWerStoLeiGorSav-CP-06} and adding long-range wave-function correlation~\cite{FroKneJen-JCP-13,HedHeiKneFroJen-JCP-13,RebTou-JCP-16}. Third, time-propagation TDRSH could be implemented to go beyond linear response and tackle strong-field phenomena, such as high-harmonic generation and above-threshold ionization~\cite{LabZapCocVenTouCaiTaiLup-JCTC-18}.



\begin{thebibliography}{80}
\bibitem[{Zap()}]{ZapLupTou-JJJ-XX-note}
\bibinfo{note}{Contrary to our Fig. \ref{Hespectra}a, the TDLDA spectrum of the
  He atom shown in Fig. 6 of Ref.~\onlinecite{WasMaiBur-PRL-03} has a larger
  maximum than the LDA spectrum. This discrepancy is due to the fact that the
  TDLDA spectrum shown in Ref.~\onlinecite{WasMaiBur-PRL-03} comes in fact from
  Ref.~\onlinecite{SteDecGor-JCP-01} where it was calculated by replacing the
  LDA 1s orbital energy by the opposite of the exact ionization energy. We have
  checked that this results not only in an energy shift of the spectrum but
  also to larger oscillator strengths. The true TDLDA spectrum of the He atom
  is thus the one shown in the present Fig. \ref{Hespectra}a.}
\end{thebibliography}
\end{document}